\documentclass[preprint,12pt]{elsarticle}
\usepackage{amssymb}
\usepackage{amsmath}
\usepackage{graphicx}
\usepackage{booktabs, makecell, tabularx}
\usepackage{braket}
\usepackage{color} 

\def\be{\begin{equation}}

\def\ee{\end{equation}}
\def\bea{\begin{eqnarray}}
\def\eea{\end{eqnarray}}
\def\bdm{\begin{displaymath}}
\def\edm{\end{displaymath}}
\journal{Nuclear Physics A}

\begin{document}

\begin{frontmatter}

\title{Nuclear Structure Properties of even-even Chromium Isotopes and
the Effect of Deformation on Calculated Electron Capture Cross Sections}

\author{Jameel-Un Nabi$^{1}$}
\author{Mahmut B\"{o}y\"{u}kata$^{2}$}
\author{Asim Ullah$^{1}$}
\author{Muhammad Riaz$^{1}$}
\address{
$^1$Faculty of Engineering Sciences, GIK Institute of Engineering
Sciences and Technology, Topi 23640, Swabi, Khyber Pakhtunkhwa,
Pakistan}

\address{
$^2$Physics Department, Science and Arts Faculty, K\i r\i kkale
University, 71450, K\i r\i kkale, Turkey}


\begin{abstract}

In this study, we investigate the role of the nuclear deformation on
the calculated electron capture cross section (ECC) of even-even
chromium (Cr) isotopes. We first determined the nuclear structure
properties of these nuclei within the interacting boson model-1 (IBM-1).
The energy spectra and E2 transition probabilities were calculated
by fitting the parameters in the model formalism. The analysis of
the potential energy surface were also performed to predict
the geometric shape of the Cr nuclei by plotting their contour plot in
the plane of ($\beta$, $\gamma$) deformation parameters. Later, we calculated
the ECC within the proton-neutron quasiparticle random phase approximation
(pn-QRPA) model. In particular we studied how the calculated ECC
changed with different values of nuclear deformation parameter.
The calculated Gamow-Teller (GT) strength distributions were widely
spread among the daughter states. The total GT strength decreased with
increasing value of the $\beta$ parameter. The computed ECC values,
however, increased with increase in the $\beta$ value of the Cr isotopes.

\end{abstract}


\end{frontmatter}

\section{Introduction}

In study of nucleosynthesis and related astrophysical phenomena, electron
capture (EC) on nuclei plays a key role. During the later phases of stellar
evolution, the electron capture cross section (ECC) has a significant impact
on reducing the ratio of electron-to-baryon content of the stellar matter
and estimating the presupernova stellar core composition as well as in
synthesizing neutron-rich (massive) nuclei \cite{bet79, bet90}. The EC
process reduces the electron degenerate pressure and eventually causes
the gravitational collapse of the core of massive star.

Study of $\beta$-decay and EC has a direct relation with investigation of
Gamow-Teller (GT) transitions and nuclear half-lives. At the beginning of
the collapse i.e at the low stellar densities ($\sim 10^{10}~gcm^{-3}$) and
the low temperatures ($0.3-0.8$~MeV), when the $Q$ value and chemical
potential of electrons have comparable magnitudes, the EC rates are responsive
to associated GT strength distributions. For higher values of temperature and
densities, the chemical potential surpasses the $Q$ value and then the EC rates
are largely dictated by the total GT strength. Therefore the computation of the
EC rates and GT strength distributions in the stellar matter is an essential
requirement. Under laboratory conditions GT transitions can be probed by
charge-changing transition reactions [e.g. ($p, n$) and ($^2He, t$)]. To study
GT transitions of  unstable nuclei  one requires a nuclear model, preferably
microscopic in nature, that can provide reliable estimate of corresponding
nuclear matrix elements in accordance with available experimental data. The
proton-neutron quasiparticle random phase approximation (pn-QRPA)~\cite{Hal67}
is one such successful  model frequently employed to  calculate the weak rates
under terrestrial \cite{Sta90,Hir93} and stellar conditions \cite{Nab99,Nab04}.

Fuller, Fowler, and Newman \cite{FFN} performed the pioneering calculation
by tabulating the weak interaction rates employing the independent-particle
model (IPM) with the help of available experimental data for astrophysical
applications. Weak rates were tabulated for nuclei with mass in the range
60 $\ge$ A $\ge$ 21. Later, the  large scale shell model (LSSM) diagonalization
model (e.g. \cite{Lan00}) and shell model Monte Carlo (SMMC) (e.g. \cite{Dea98})
were  utilized to improve $\beta^{\pm}$-decay, positron and EC rates with the help
of computed GT strength distributions for various nuclei in mass range
45 $\le$ A $\le$ 65. Based on these revised weak interaction rates, the pre-supernova
phases of heavy mass stars were investigated in Ref.~\cite{Lan01}. A detailed
tabulation of stellar rates was also provided by SMMC and pn-QRPA model in
Refs.~\cite{Nab04,Dea98}. It was concluded that EC rates on the $fp$-shell nuclei
and their GT strength distributions significantly effect the presupernova evolution
of heavy-mass stars. Recently, ECC and  EC rates were computed using the pn-QRPA
model for various $fp$-shell nuclei~\cite{Nab19}. These stellar EC rates were found
bigger than the former calculations when compared at high stellar temperatures.
At the same time, the computed ECC were noted smaller in contrast to the shell model
(SM) results at the low electron incident energies.

For investigation of nuclear structure, one of the useful models is the
interacting boson model-1 (IBM-1)~\cite{Iachello87} established on Lie algebra
taking into account the U(6) group structure of interacting $s$ and $d$ bosons
corresponding to the angular momentum $L=0$ and $L=2$, respectively. This model
is fairly successful for the description of the nuclear structure for even-even
nuclei, in particular for medium- and heavy-mass ones except those having magic
numbers along the nuclear chart~\cite{Cas88}. Some nuclear structure properties
such as the energy levels, the reduced electromagnetic transition probabilities
$B(E2)\downarrow$, and the prediction of the nuclear geometry for A$\sim$100
region were reported in Refs.~\cite{Boy08,Boy10i,Boy10ii,Boy14}.

During the last two decades some crucial studies on neutron-rich Cr nuclei were
performed in both experimental and theoretical areas. A brief review of the main
findings are presented as follows for even-even Cr isotopes. Experimental $2^+$
levels of $^{60-62}$Cr isotopes were found through the $\beta$-delayed neutron decay of
$^{61-63}$V isotopes and their deformations were claimed in Ref.~\cite{Sor03}.
The expanded level schemes of $^{56,58,60}$Cr were experimentally investigated
at the Gammasphere using two different types of reactions, the results compared
with the SM calculation and their ground-state energy surfaces were presented
from total Routhian surface (TRS) calculations~\cite{Zhu08}. The structure of
$^{52-62}$Cr isotopes was systematically investigated by Kaneko
$et. al.$~\cite{Kan08}, the known energy levels of these isotopes and their
B(E2) values were compared and also unknown levels of $^{62,64,66}$Cr were
predicted within the spherical SM calculations. The collectivity of even-even
N=40 isotones were investigated using the Hartree-Fock-Bogoliubov (HFB)
approach~\cite{Gau09}, some unknown energy levels and B(E2) values of $^{64}$Cr
were predicted and its shape was presented as spherical by plotting triaxial
potential energy surfaces. The collectivity of $^{60-64}$Cr with neighboring
Fe isotopes, around N$=$40 neutron-rich region, were investigated by comparing
the experimental excitation energies with the LSSM calculation and the excited
states of $^{64}$Cr isotope were measured ~\cite{Gad10}. The projected SM
calculation was performed on the moments of inertia of yrast bands for
$^{54-68}$Cr by comparing the experimental data \cite{Yan10}. The collective
Hamiltonian approach was applied to study the shape-phase transition of
$^{58-68}$Cr isotopes and $^{60}$Cr was reported close to the critical point~\cite{Yos11}.
The excited states and the B(E2) values of $^{56}$Cr were measured and the
results were compared with the LSSM calculation~\cite{Sei11}. The coulomb
excitation measurements were performed on $^{58-62}$Cr isotopes and B(E2)$\uparrow$
rates of $^{60,62}$Cr were estimated by deduction from the measured cross sections~\cite{Bau12}.
The quadrupole collectivity of the low-lying states for $^{58-64}$Cr isotopes were
investigated for the shape transition along given isotopic chain by
using the constrained HFB and the local QRPA methods~\cite{Sat12}. The B(E2)
values of $^{64}$Cr was determined by performing intermediate-energy Coulomb
excitation measurements in Ref.~\cite{Cra13}. The collective features of
$^{54-64}$Cr isotopes and their neighbors Fe isotopes were investigated within
the interacting boson model-2 (IBM-2) and the interacting shell model  for their shape
transition and it was found that $^{58}$Cr was located around the critical
point~\cite{Kot14}. The low-lying energy states, B(E2) values of $^{54-66}$Cr
and $^{56-68}$Fe isotopes were calculated by performing the angular-momentum
projection on the mean-field energy surface~\cite{Jia14}. The $2^+_{1}$ and
$4^+_{1}$ states of $^{66}$Cr and $^{70,72}$Fe isotopes were measured by
in-beam $\gamma$-ray spectroscopy~\cite{San15} and also their quadrupole
deformation properties were calculated within the LSSM model. Within the same
model, the energy levels and the total energy surfaces of $^{56,58}$Cr with
neighbor were investigated by Togashi $et. al.$~\cite{Tog15}. Few B(E2) values
of $^{58-62}$Cr were obtained by measuring the lifespans of their low-lying
states and later compared within IBM-2 and SM calculations in Ref.~\cite{Bra15}.
The $2^+_{1}$ states and new B(E2) values of $^{48,50}$Cr with neighbor nuclei
along the N$=$Z line were measured and compared with SM calculation~\cite{Arn17,Arn18}.

Recently, pn-QRPA and IBM-1 models were used to describe the structure of some
even-even nuclei positioned along the $Z=N$ line~\cite{Nabi16,Nabi17a,Nabi17b,Nabi19,Ull20}.
More recent studies include the role of the nuclear deformation on the calculated  GT
strength distributions and ECC on $^{46,48,50}$Cr isotopes within the
formalism of the pn-QRPA model~\cite{Nabi19,Ull20}.
Moreover, the deformation effects on the cross sections were also investigated
for the $^{110}$Pd(d,n)$^{111}$Ag and $^{110}$Pd(d,2n)$^{110m}$Ag reactions~\cite{Boy18}.
For the current work, we aim to study in detail  the nuclear structure
properties of even-even Cr nuclei along the $Z=28$ isotopic chain and
explore the effect of nuclear deformation on the calculation of ECC rates.
The IBM-1 was applied to calculate the energy levels and B(E2)
values of the selected Cr isotopes  by fitting Hamiltonian
parameters. Their geometric shapes were also predicted within the energy
surface formalism including common parameters of the model Hamiltonian.
These energy surfaces were plotted to determine the $\beta$ deformation
parameters and these deformations were also calculated using the experimental
$B(E2)\uparrow$ values. Later, the pn-QRPA model was employed for the
computation of ECC values on given Cr isotopes. The ECC values were calculated
by using varying deformation parameters  to
study the possible correlation between calculated ECC and deformations.

This paper is structured as follows: some recent studies on the given Cr nuclei
were briefly reviewed in the introduction part. In Section~2 we briefly introduce
formalism for both IBM-1 and pn-QRPA models. The results are
discussed and compared with experimental and previous theoretical calculations
in Section~3. Finally, the concluding remarks are stated in the last section.

\section{Model Formalism}
\subsection{Interacting Boson Model-1 (IBM-1)}
\label{ss_ibm}
The IBM-1~\cite{Iachello87} is a very useful group theoretical model to describe
the nuclear structural properties of  even-even nuclei. This Lie
algebraic model is established on the unitary group $U(6)$~\cite{Iachello06}.
The model Hamiltonian has several forms~\cite{Iachello87} and for the present
application, we used following form including the spherical and deformed parts:
\begin{equation}
\hat H= \epsilon_d\,\hat n_d + a_1 \,\hat Q\cdot\hat Q + a_2\,\hat L\cdot\hat L , \label{ham}
\end{equation}
where $\hat n_d$ (= $\sqrt{5}[d^\dag\times\tilde d]^{(0)}$) represents the
boson-number, $\hat L$ (=$\sqrt{10}[d^\dag\times\tilde d]^{(0)}$) is the
angular momentum operator and $\hat Q$ is quadrupole operator given by
$\hat Q=[d^\dag\times\tilde s+s^\dag\times\tilde d]^{(2)}+\overline{\chi}[d^\dag\times\tilde d]^{(2)}$.
The constants $\epsilon_d$, $a_1$ and $a_2$ in Eq. (\ref{ham}) and
$\overline{\chi}$ in the quadrupole operator $\hat Q$ are free parameters
fitted by calculating energy levels and comparing them with experimental data obtained from National Nuclear Data Center (NNDC)~\cite{NNDC16}.

The IBM-1 Hamiltonian can be used to obtain the energy surface of a given
nucleus in the classical limit which can be helpful in predicting the geometry
of the nucleus~\cite{Dieperink80,Dieperink80b,Ginocchio80,Ginocchio80b,Isacker81}.
This energy surface can be written in terms of the $\beta$ and $\gamma$
deformation parameters as following:

\begin{equation}\label{pes}
\begin{split}
V(\beta,\gamma)=\epsilon_d \frac {N \beta^2}{1+\beta^2} + a_2 \frac{6 N \beta^2}{1+\beta^2} + a_1 \frac{ N }{1+\beta^2}
~~~~~~~~~~~~~~~~~~~~~~~~~~~~~~~~~~~~~~~\\
 \times \left[ 5 + (1 + \overline{\chi}^2) \beta^2 +
\frac{(N-1)\left(\frac{2\overline{\chi}^2\beta^2}{7}-4\sqrt{\frac{2}{7}}\overline{\chi}\beta\cos3\gamma+4\right)\beta^2}{1+\beta^2}\right]
\end{split}
\end{equation}
including common parameters $\epsilon_d$, $a_1$, $a_2$ and
$\overline{\chi}$ in  Eq.~(\ref{ham}).
The $\beta$ and $\gamma$ parameters play same role of the geometric model
of Bohr and Mottelson~\cite{Bohr98}. These parameters evaluate the axial
drift from the spherical to the deformed side and the angle shift from
the axial deformation, respectively. It is possible to determine these
deformation parameters by minimizing the potential energy given in Eq.~(\ref{pes}).

The $\beta$ parameter may also be obtained from the reduced electric
quadrupole transition rate ($B(E2)\uparrow$) for the transition from
the ground state ($0^+_1$) to the first excited state ($2^+_1$). This
relationship of the $\beta$ and the adopted $B(E2)\uparrow$ value is
given as follows ~\cite{ram01}:
\begin{equation}
\beta=\left(\frac{4\pi}{3 Z R^2_0}\right) \left[\frac{B(E2)\uparrow}{e^2}\right]^{1/2}, \label{def2}
\end{equation}
where $Z$ is stands for proton number, $R_0$ represents the nuclear
radius given by
\begin{equation}
R^2_0=(1.2\times 10^{-13} R^{1/3} cm)^2=0.0144 A^{2/3}b. \label{R0}
\end{equation}
The $B(E2)\uparrow$ value has a relation to $B(E2)\downarrow$ value
and can be obtained by
\begin{equation}
 B(E2)\uparrow =\left(\frac{2J_f + 1}{2J_i + 1}\right) B(E2)\downarrow, \label{BE2up}
\end{equation}
 here $B(E2)\uparrow$ value is B(E2:$0^+_{1}\rightarrow2^+_{1}$) and
 $B(E2)\downarrow$ value refers to B(E2:$2^+_{1}\rightarrow0^+_{1}$)
 transitions. Both values may be obtained from the Evaluated Nuclear
 Structure Data File (ENSDF) of the NNDC~\cite{NNDC16}
 and may be converted from one to the other.

The B(E2) values of selected nuclei may further be calculated within IBM-1
model~\cite{Iachello87} by using the reduced matrix elements of $E2$
given by $\hat T(E2)=e_b \hat Q$. Here, $\hat Q$ is the common
quadrupole operator given in Eq.~(\ref{ham}) and the constant
$e_b$ is the boson effective charge used as a free parameter.
\subsection{The proton-neutron quasiparticle random phase approximation (pn-QRPA)}
The following pn-QRPA Hamiltonian was considered for computing the GT
strength distributions and associated ECC on the selected Cr isotopes;
\begin{equation} \label{H}
H^{QRPA} = H^{sp} + V^{pp}_{GT} + V^{ph}_{GT} + V^{pair},
\end{equation}
where $H^{sp}$ denotes the single particle Hamiltonian, $V_{GT}^{pp}$
and $V_{GT}^{ph}$ terms are the particle-particle and particle-hole GT
forces, respectively. Single particle energies and wave
functions were calculated in the Nilsson model \cite{nil55}, in
which the nuclear deformation was considered. The last term $V^{pair}$ represent the pairing
force for which the BCS approximation was considered. For the complete recipe to diagonalize Eq.~(\ref{H}) we refer to \cite{Nab19}. We briefly explain the formalism below.

In the current work, we began with a spherical nucleon basis $(c^{\dagger}_{jm}, c_{jm})$, having j as total angular momentum with z-component $m$ which was later transformed to a deformed (axial-symmetric) basis denoted by $(d^{\dagger}_{m\alpha}, d_{m\alpha})$. The transformation matrix elements were given by

\begin{equation}\label{df}
	d^{\dagger}_{m\alpha}=\Sigma_{j}D^{m\alpha}_{j}c^{\dagger}_{jm}.
\end{equation}
where $D$ is the transformation matrix consisting of a set of Nilsson eigenfunctions, and $\alpha$ represents additional quantum numbers. The BCS
calculation for the neutron and proton systems was performed separately. This reduced the QRPA matrix to an algebraic equation of fourth order. This problem was much easier to solve as compared to full diagonalization of the non-Hermitian matrix of
large dimensionality. We adopted a constant pairing force of strength G ($G_p$ and $G_n$ for neutrons and protons, respectively),
\begin{equation}\label{pr}
	\begin{split}
		V_{pair}=-G\sum_{jmj^{'}m^{'}}(-1)^{l+j-m}c^{\dagger}_{jm}c^{\dagger}_{j-m}\\
		(-1)^{l^{'}+j^{'}-m^{'}} c_{j^{'}-m^{'}}c_{j^{'}m^{'}},
	\end{split}
\end{equation}
the summation over $m$ and $m^{'}$ was limited to $m$, $m^{'}$ $>$ 0, and $l$ is the orbital angular momentum. From BCS calculation, we computed the occupation amplitudes $u_{m\alpha}$ and $v_{m\alpha}$ (satisfying $u^{2}_{m\alpha}$+$v^{2}_{m\alpha}$=1) and quasiparticle (q.p.) energies $\varepsilon_{m\alpha}$. A q.p. basis $(a^{\dagger}_{m\alpha}, a_{m\alpha})$ was later introduced from the Bogoliubov transformation
\begin{equation}\label{qbas}
	a^{\dagger}_{m\alpha}=u_{m\alpha}d^{\dagger}_{m\alpha}-v_{m\alpha}d_{\bar{m}\alpha},
\end{equation}

\begin{equation}
a^{\dagger}_{\bar{m}\alpha}=u_{m\alpha}d^{\dagger}_{\bar{m}\alpha}+v_{m\alpha}d_{m\alpha},
\end{equation}
where $\bar{m}$ is the time reversed state of $m$ and $a^{\dagger} (a)$ is the q.p. creation (annihilation) operator which comes in the RPA equation. Creation operators of QRPA phonons can be represented by
\begin{equation}\label{co}
	A^{\dagger}_{\omega}(\mu)=\sum_{pn}[X^{pn}_{\omega}(\mu)a^{\dagger}_{p}a^{\dagger}_{\overline{n}}-Y^{pn}_{\omega}(\mu)a_{n}a_{\overline{p}}].
\end{equation}
In the above equation, the indices $p$ and $n$ stand for $m_{p}\alpha_{p}$ and
$m_{n}\alpha_{n}$, respectively, and distinguish between proton and neutron single-particle states. The sum was taken over proton-neutron pairs satisfying $\mu=m_{p}-m_{n}$ and $\pi_{p}.\pi_{n}$=1, with $\pi$ denoting parity. In our model, proton-neutron residual interactions occurred as particle-hole ($\chi$) and particle-particle ($\kappa$) interactions. The $ph$ GT force was then represented by
\begin{equation}\label{ph}
	V^{ph}= +2\chi\sum^{1}_{\mu= -1}(-1)^{\mu}Y_{\mu}Y^{\dagger}_{-\mu}\\
\end{equation}
\begin{equation}\label{ph}
	Y_{\mu}= \sum_{j_{p}m_{p}j_{n}m_{n}}<j_{p}m_{p}\mid
	t_- ~\sigma_{\mu}\mid
	j_{n}m_{n}>c^{\dagger}_{j_{p}m_{p}}c_{j_{n}m_{n}},
\end{equation}
and the $pp$ GT force as
\begin{equation}\label{ph}
	V^{pp}= -2\kappa\sum^{1}_{\mu=
		-1}(-1)^{\mu}P^{\dagger}_{\mu}P_{-\mu}.
\end{equation}
with
\begin{equation}\label{p}
P^{\dagger}_{\mu}= \sum_{j_{p}m_{p}j_{n}m_{n}}<j_{n}m_{n}\mid
(t_- \sigma_{\mu})^{\dagger}\mid
j_{p}m_{p}>(-1)^{l_{n}+j_{n}-m_{n}}c^{\dagger}_{j_{p}m_{p}}c^{\dagger}_{j_{n}-m_{n}},
\end{equation}
Here, the $ph(pp)$ force is specified with a positive (negative)
sign, since the $ph(pp)$ force in $J^{\pi}=1^{+}$ channel is generally repulsive (attractive), and then the interaction strength, $\chi$ and $\kappa$, takes positive values. The selections of these
two constants were done in an optimal fashion to reproduce available
experimental data and fulfillment of model independent Ikeda sum rule~~\cite{Ikeda1963}. In this paper, we chose the value of $\chi$ to be
4.2/A, showing a $1/A$ dependence \cite{Hom96} and $\kappa$ equal to
0.10. See also Ref.~\cite{Maj16}.
In RPA equation matrix elements of the forces are separable,
\begin{equation}\label{phs}
	V^{ph}_{pn,p^{'}n^{'}}= +2\chi f_{pn}(\mu)f_{p^{'}n^{'}(\mu)},\\
\end{equation}
\begin{equation}\label{pps}
	V^{pp}_{pn,p^{'}n^{'}}= -2\kappa f_{pn}(\mu)f_{p^{'}n^{'}(\mu)},
\end{equation}
with
\begin{equation}\label{f}
	f_{pn}(\mu)=\sum_{j_{p}j_{n}}D^{m_{p}\alpha_{p}}_{j_{p}}D^{m_{n}\alpha_{n}}_{j_{n}}<j_{p}m_{p}\mid
	t_-~\sigma_{\mu}\mid j_{n}m_{n}>,
\end{equation}
which are single-particle GT transition amplitudes defined in the
Nilsson basis. For the separable forces, the matrix equation can be
explicitly defined as
\begin{equation}\label{x}
	\begin{split}
		X^{pn}_{\omega}=\frac{1}{\omega-\varepsilon_{pn}}[2\chi(q_{pn}Z^{-}_{\omega}+\tilde{q_{pn}}Z^{+}_\omega)\\
		-2\kappa(q^{U}_{pn}Z^{- -}_{\omega}+q^{V}_{pn}Z^{+ +}_{\omega})],
	\end{split}
\end{equation}
\begin{equation}\label{y}
	\begin{split}
		Y^{pn}_{\omega}=\frac{1}{\omega+\varepsilon_{pn}}[2\chi(q_{pn}Z^{+}_{\omega}+\tilde{q_{pn}}Z^{-}_\omega)\\
		+2\kappa(q^{U}_{pn}Z^{+ +}_{\omega}+q^{V}_{pn}Z^{- -}_{\omega})],
	\end{split}
\end{equation}
where $\varepsilon_{pn}=\varepsilon_{p}+\varepsilon_{n}$,\\ \\
$q_{pn}=f_{pn}u_pv_n$, $q_{pn}^{U}=f_{pn}u_pu_n$,\\ \\
$\tilde q_{pn}=f_{pn}v_pu_n$, $q_{pn}^{V}=f_{pn}v_pv_n$,\\ \\
\begin{equation}\label{Z-}
	Z^{-}_{\omega}= \sum_{pn}(X^{pn}_{\omega}q_{pn}-Y^{pn}_{\omega}\tilde q_{pn})\\
\end{equation}
\begin{equation}\label{Z+}
	Z^{+}_{\omega}= \sum_{pn}(X^{pn}_{\omega}\tilde q_{pn}-Y^{pn}_{\omega}q_{pn})\\
\end{equation}
\begin{equation}\label{Z--}
	Z^{- -}_{\omega}= \sum_{pn}(X^{pn}_{\omega}q^{U}_{pn}+Y^{pn}_{\omega}q^{V}_{pn})\\
\end{equation}
\begin{equation}\label{Z++}
	Z^{+ +}_{\omega}=
	\sum_{pn}(X^{pn}_{\omega}q^{V}_{pn}+Y^{pn}_{\omega}q^{U}_{pn}).
\end{equation}
Ultimately, we were left with a matrix equation,
\begin{equation}\label{M}
	M z=0,
\end{equation}
with solution given by
\begin{equation}\label{DM0}
	det M=0,
\end{equation}
Consequently we reduced the solution of conventional  eigenvalue problem of
the RPA equation to determining roots of algebraic equation (Eq.~\ref{DM0}). For details of solution of Eq.~\ref{DM0} we refer to \cite{mut92}. Applying normalization condition to phonon amplitudes
\begin{equation}\label{DM1}
	\sum_{pn}[(X^{pn}_{\omega})^{2}-(Y^{pn}_{\omega})^{2}]=1,
\end{equation}
the absolute values were found by substituting $Z_{\omega}'s$ into
Eq.~\ref{x} and Eq.~\ref{y}.

Our calculated GT strength distributions fulfilled the Ikeda sum rule (ISR)~\cite{Ikeda1963}. We used $\hbar\omega=41A^{1/3}$ to compute the oscillator constant for nucleons. The Nilson-potential parameters were taken from Ref.~ \cite{ragnarson1984}. The pairing gaps were computed using   $\vartriangle_p=\vartriangle_n={12/\sqrt A}$  MeV~\cite{hardy09}.
$Q$-values were adopted from Ref. \cite{audi2017}. We discuss the deformation parameter in next section. 
The electron capture (EC) weak rate from
parent state $|m\rangle$ to daughter state $|n\rangle$ is given by
\begin{equation}\label{ec}
	\lambda_{mn}^{EC}= \ln2 \dfrac{f_{mn}^{EC}(T, \rho, E_f)}{D/B_{mn}},
\end{equation}
where $f_{mn}^{EC}(T, \rho, E_f)$ are the phase space integrals. For details of solution of these integrals we refer to~\cite{Nab99, Nab04}.
$B_{mn}$ is the nuclear reduced transition probability and
is given by
\begin{equation}\label{5}
B_{mn}=(g_A/g_V )^2 B(GT)_{mn} + B(F)_{mn}.
\end{equation}
We took $D$ = 6143$s$ and $g_A/g_V$ =  -1.254 from Refs. \cite{Nak10}
and ~\cite{hardy09}, respectively. The reduced Fermi $B(F)_{mn}$ and
GT $B(GT)_{mn}$ transition probabilities  were computed using the
following relations
\begin{equation}
B(F)_{mn}=\dfrac{1}{2J_m+1} |\langle n\|\sum_kt_+^k \|m\rangle|^2
\end{equation}
\begin{equation}
B(GT)_{mn}=\dfrac{1}{2J_m+1} |\langle n\|\sum_kt_+^k\vec{\sigma}^{k} \|m\rangle|^2,
\end{equation}
where $J_m$ represents the total spin of the parent state $|m \rangle$,
$\vec{\sigma}^{k}$ are the Pauli spin matrices and $t_+^i$ refer
to the iso-spin raising operator. The weak-interaction Hamiltonian which
governs the calculation of electron capture cross-section is given by
\begin{equation}
\widehat{H}_{\omega}=\dfrac{G}{\sqrt2}j_\mu^{lept}\widehat{J}^{\mu},
\end{equation}
where G=$G_Fcos\theta_c$. The notations $\theta_c$ and $G_F$ stand for Cabibbo
angle and Fermi coupling constant, respectively. The leptonic current
$j_\mu^{lept}$ and hadronic current $\widehat{J}^{\mu}$ are given by
\begin{equation}
j_{\mu}^{lep}=\bar{\psi}_{\upsilon_{e}}(x)\gamma_{\mu}(1-\gamma_5)\psi_{\upsilon_{e}}(x)
\end{equation}

\begin{equation}
\hat{J}^{\mu}=\bar{\psi}_{p}(x)\gamma_{\mu}(1-c_{A}\gamma_5)\psi_{n}(x),
\end{equation}
where the $\psi_{\nu_e}(x)$ are spinor operators and other symbols have usual meaning.
The computation of ECC is based on the matrix elements between initial
state $\Ket{m}$ and final state $\ket{n}$ of parent and daughter nuclei,
respectively.
\begin{equation}
\bra{n}|\widehat{H}_{\omega}|\ket{m}=\dfrac{G}{\sqrt2}l^{\mu}\int
d^3xe^{-i{q.x}}\bra{n}\hat{J_{\mu}}\ket{m},
\end{equation}
The term $q$ in the above equation refers to the three-momentum transfer
and $l^\mu e^{-iq.x}$ stands for the leptonic matrix element which was
employed in matrix element calculation \cite{NPaar2009, walecka2004}.
We assumed the low momentum transfer approximation $q \longrightarrow 0$
in this work. This approximation justified the GT
operator ($GT^+ =\sum_{m}\tau_m^+\sigma_m$) as the  dominant contributor to the total ECC~\cite{walecka2004}.
The total ECC in terms of incident electron energy ($E_e$) may then be
computed by using following equation:
\begin{equation}\label{CS}
\begin{split}
\sigma_c(E_e,T)=\dfrac{G_F^{2}cos^2\theta_c}{2\pi}\sum\limits_{m}F(E_e,Z)\dfrac{(2J_{m}+1)\exp{(-E_m/kT)}}{G(Z,A,T)}\\
\times
\sum\limits_{J,f}(E_e+E_m-E_n-Q)^{2}\dfrac{|\bra{m} GT^{+} \ket{n}|^2}{(2J_{m}+1)}.
\end{split}
\end{equation}
The Fermi function represented by $F(E_e, Z)$ in the above equation was
computed using the recipe of Ref.~\cite{gov71}. To compute the
nuclear partition function $G(A, Z, T)$, we used the prescription recently
introduced in Refs. \cite{nabo2016,nabi2016}. This recipe is believed to give a more realistic estimate of the partition functions appearing in Eq.~(\ref{CS}). Other symbols have usual meaning.

\section{Results and Discussions}

For the nuclear structure investigation of given isotopes,  we first
calculated the  energy levels within the IBM-1 Hamiltonian given by Eq.
(\ref{ham}) by fitting the parameters $\epsilon_d$, $a_1$, $a_2$ and
$\overline{\chi}$ listed in Table-\ref{par} for even-even $^{46-50,54-66}$Cr
along $Z=24$ isotopic chain (except for $^{52}$Cr since its neutron number is
magic). For the fitting procedure, we tried to minimize the \emph{root mean
square} ($rms$) deviations, in the energies and $B$(E2) values, donated 
by $\sigma$ as defined in following formula

\begin{eqnarray}\label{rms}
\sigma&=&\sqrt{\frac{1}{N}\sum_i\left(exp.^i-cal.^i\right)^2},
\end{eqnarray}
where the sums are over the available experimental data and calculated 
ones by IBM-1, $N$ is the number of the experimental data for the 
energy levels and the B(E2) values. The minimized $\sigma$ 
values are listed in the last column of Table-\ref{par} and Table-\ref{be2} for the energy levels and B(E2), respectively.

\begin{figure}
\begin{center}
\includegraphics[width=14cm]{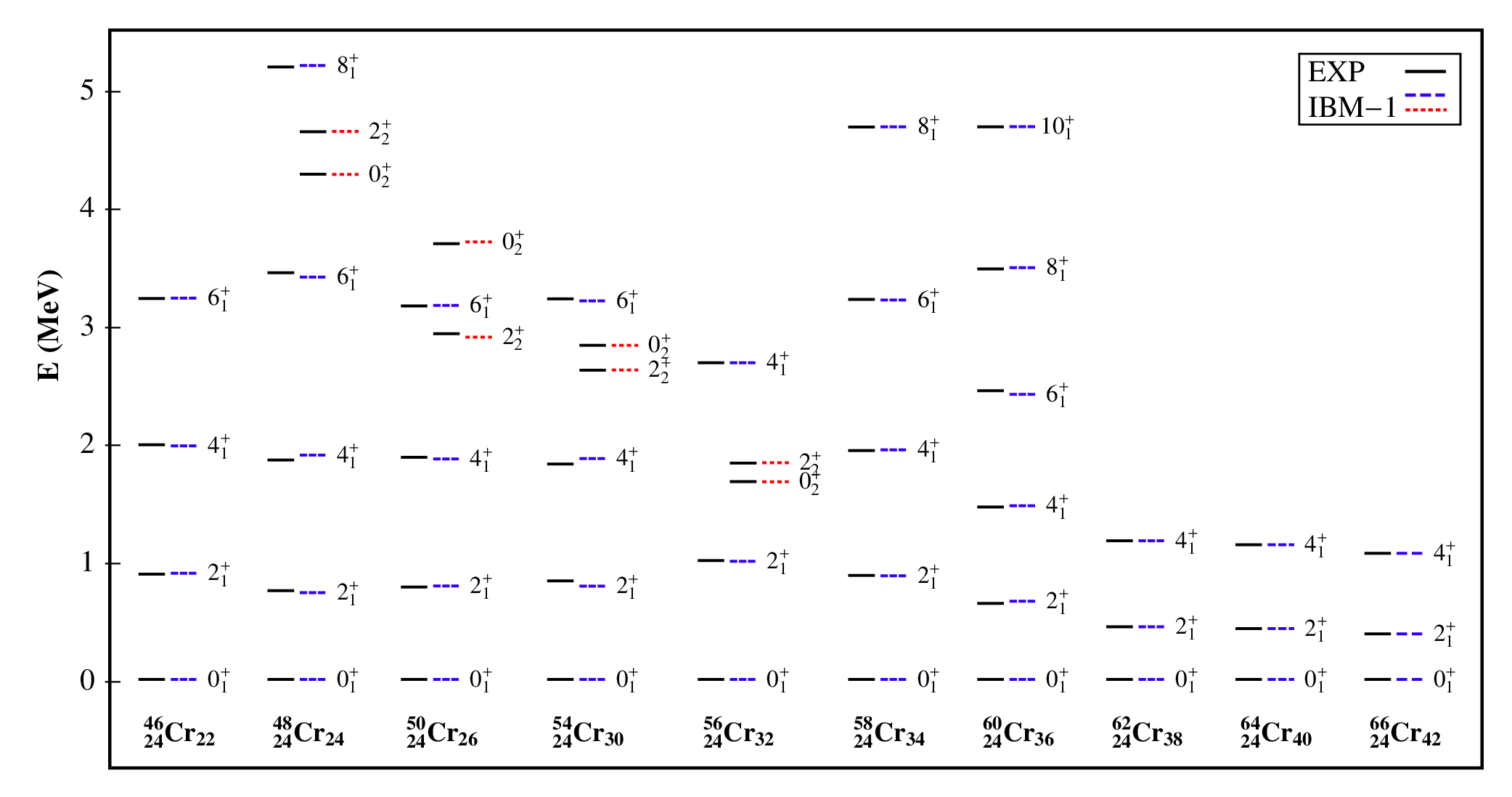}
\end{center}
\caption{(Color online) The experimental \cite{NNDC16}
and calculated energy levels of selected Cr isotopes. Dashed (blue) lines show
the levels in the ground-state bands and dotted (red) ones are the other levels.} \label{f_en}
\end{figure}

The calculated energy levels of selected Cr isotopes are shown in Figure~(\ref{f_en})
with experimental data obtained from NNDC~\cite{NNDC16}. In this figure, the
experimental data are denoted as solid line and the calculated ones are shown
as dashed and dotted lines. The known second $0^{+}$ and $2^{+}$ levels in beta
bands of $^{~48,~50,~54,~56}$Cr isotopes are also calculated with IBM-1 and they
are in good agreement with experimental data.

The computed and experimental B(E2) values~\cite{NNDC16} of Cr isotopes are listed
in Table-\ref{be2}. Here, the boson effective charge ($e_b$) which is a constant in
the reduced matrix elements of $E2$ given by $\hat T(E2)=e_b \hat Q$ is fitted by
minimizing the $rms$ deviations ($\sigma$) for all obtained B(E2) values.

\begin{figure}
\begin{center}
\includegraphics[width=15cm]{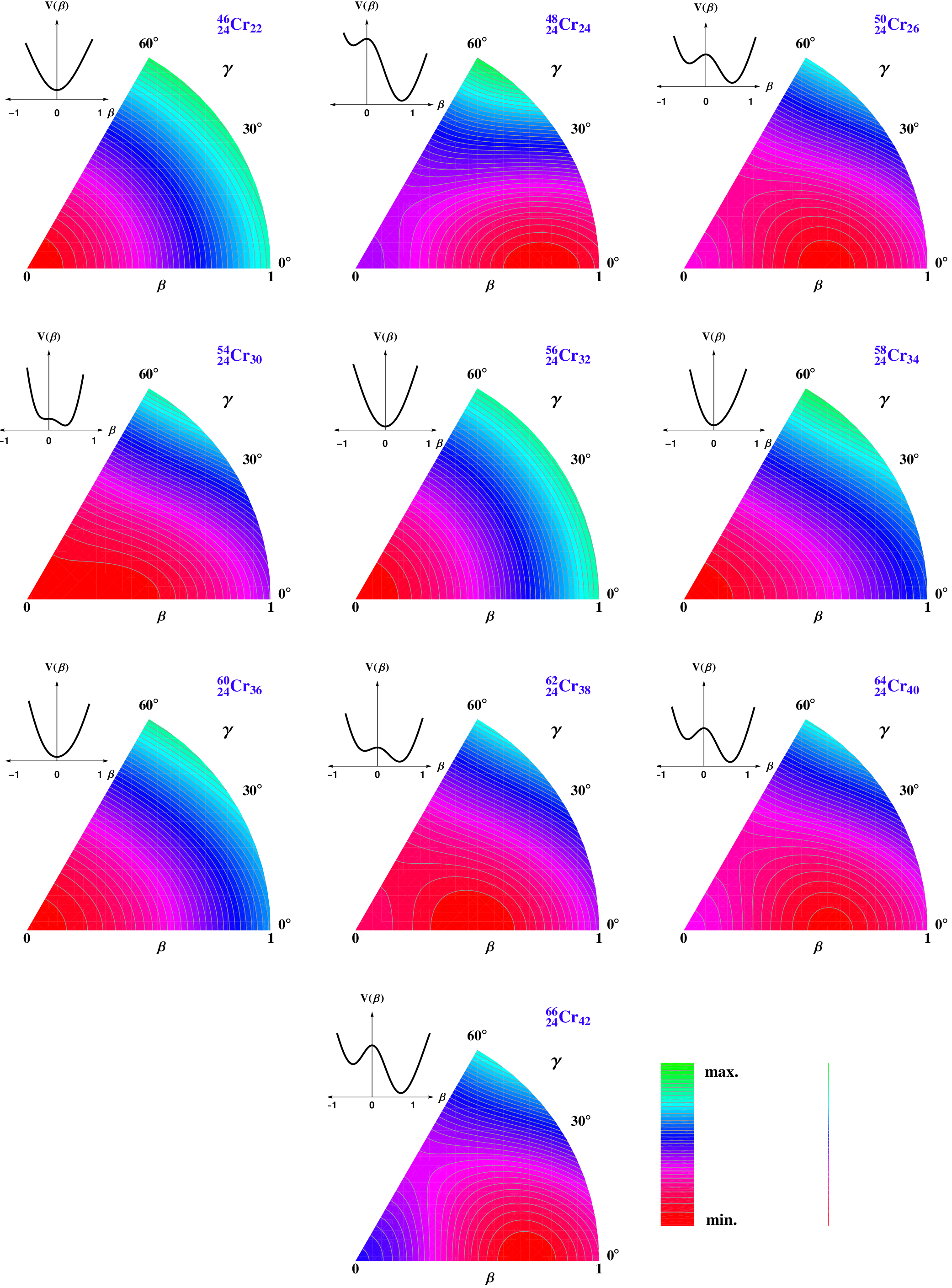}
\end{center}
\caption{(Color online) The contour plot of the energy surfaces in the ($\beta$, $\gamma$)
plane and as a function of $\beta$ for $\gamma$ = $0$ for the selected Cr isotopes.} \label{f_pes}
\end{figure}

For calculation of the deformation parameters listed in Table-\ref{A}, we performed the
following procedures. First we plotted the contour plot of the potential energy
surface of IBM-1 as function of deformation parameters $(\beta,\gamma)$ given in Eq.
(\ref{pes}) for each selected Cr isotopes as seen in Figure~(\ref{f_pes}). This figure
also includes the energy surfaces as a function of $\beta$ for $\gamma=0$. According
to these contour plots, $^{46}$Cr, $^{56}$Cr, $^{58}$Cr, and $^{60}$Cr are
spherical since their deformation parameters are \emph{zero} while $^{48}$Cr, $^{50}$Cr,
$^{54}$Cr, $^{62}$Cr, $^{64}$Cr, and $^{66}$Cr  have axially deformed shapes.
The calculated deformation parameters obtained by the IBM-1 are listed in Table-\ref{A}
and denoted as  $\beta_{IBM-1}$. The experimental deformation parameter of Cr
isotopes was calculated using the experimentally adopted $B(E2)\uparrow$ values~\cite{NNDC16}
by using the recipe given in Eqs. (\ref{def2}), (\ref{R0}), (\ref{BE2up}). These experimental deformation parameters are denoted as
$\beta_{B(E2)}$ in the same table. The $\beta_{GM}$ ($GM$ stands for geometric
model) was calculated from the $\beta$ values of the IBM-1 using the following relation
\begin{equation}
	\beta_{GM} \lesssim 1.18(2N/A)\beta_{IBM}
\end{equation}
where \emph{N} is the number of valence nucleons between the magic numbers and \emph{A}
is the mass number of the nuclei~\cite{Ginocchio80b}. Finally, Macroscopic (Yukawa-plus-exponential)-microscopic
(Folded-Yukawa) model (Mac-mic model)~\cite{muller81} was used for the calculation of the
fourth set of deformation parameters of  Cr isotopes. The electric quadrupole moment
($Q_2$) may be calculated within the Mac-mic model and was then later substituted in the following
equation to compute the deformation ($\beta$) values
\begin{equation}
\beta_{Mac-mic}=\frac{125\hspace{0.01in} Q_2}{1.44 \hspace{0.01in} Z \hspace{0.01in}A^{2/3}}.	
\end{equation}
These four different nuclear deformation parameters, denoted as $\beta_{IBM-1}$, $\beta_{B(E2)}$,
$\beta_{GM}$ and  $\beta_{Mac-mic}$ with their values listed in Table-\ref{A}, were employed later to study their effects on the calculated ECC on selected Cr isotopes. Table-\ref{A}  displays the total strength values computed from the calculated GT strength
distributions for the selected even-even Cr isotopes. The table shows that the total strength
values reduces as the mass number increases. This outcome is simply attributed to the fact that for
bigger neutron number, the EC process becomes more challenging. Table-\ref{A} highlights the effect of deformation
parameter on the computed GT strength values. It can be seen that for each  Cr isotope, with decrease in
the value of $\beta$, the calculated GT strength increases. We, however, are not in a position to generalize this
statement which warrants further investigation.

Figures~(\ref{46-56CR_GT}-\ref{58-66CR_GT}) show the skyscrapers for the pn-QRPA computed GT strength distributions along EC direction for the selected even-even Cr isotopes. The GT strengths are shown in arbitrary units, while the excitation energies in the daughter are in units of MeV. For these figures we chose  $\beta_{Mac-mic}$ values for the deformation parameter (see Table-\ref{A}). It is seen from the figures that GT strengths are fully fragmented among the daughter states of the nuclei.  The strength comparatively decreases as the mass number increases as the EC becomes more challenging. As discussed before these strength distributions fulfilled the model independent Ikeda sum rule~\cite{Ikeda1963}. 

Figures~(\ref{46-64CR_a}-\ref{46-64CR_b}) show the calculated ECC on the selected even-even Cr
isotopes ($^{46-64}$Cr), respectively, as a function of incident electron energy ($E_e$) in the range
(0--30) MeV. Each panel shows calculated ECC at three stellar temperatures shown in the legend. It maybe noted
that $^{52}$Cr and $^{66}$Cr are not shown because
only $\beta_{B(E2)}$ and $\beta_{Mac-mic}$ were calculated for $^{52}$Cr and $\beta_{B(IBM-1)}$,
$\beta_{GM}$, $\beta_{Mac-mic}$ were computed for $^{66}$Cr (see Table-\ref{A}).
From these figures, it is noted that with increasing $E_e$, the calculated  ECC increases because of the
$(E_e + E_m - E_n - Q)^{2}$ factor in Eq.~(\ref{CS}). It is also noted that there is an exponential
increase in ECC in the energy range (0--30) MeV. When the core temperature rises from 0.5 MeV to
1.0 MeV, there is a prominent increase in the calculated ECC value up to a factor 50. Configuration
mixing and thermal unblocking of states could be cited as probable sources for this increment \cite{langanke2001}.
With a further increase in temperature from 1.0 MeV to 1.5 MeV, there is very small increase (less
than a factor 2) in the  computed ECC. This is because majority of the transitions are already
unblocked at these temperatures. The steep increase may be traced to the behavior of the computed GT
strength distribution.  The centroid of the GT strength distribution shifts by few MeV with increase
in temperature. The nuclear partition function also increases with increase in core temperature. These
factors are responsible for the change in behavior of the ECC.

The effect of changing deformation on computed ECC was finally investigated. Figure~(\ref{d46-64CR})
shows the calculated ECC as a function of deformation parameter at a fixed stellar temperature of 1 MeV.
In Figure~(\ref{d46-64CR}), there are nine panels each representing even-even Cr isotope from
$^{46-64}$Cr excluding $^{52}$Cr and $^{66}$Cr for the reason mentioned above.  It may be deduced from
the figure that the computed ECC on the Cr isotopes decreases with decrease in the value
of nuclear deformation and vice versa. This is an interesting finding but as mentioned before warrants further investigation.
\section{Conclusion}

As reviewed  in the first section, some experimental and theoretical studies
~\cite{Sor03,Zhu08,Kan08,Gau09,Gad10,Yan10,Yos11,Sei11,Bau12,Sat12,Cra13,Kot14,Jia14,San15,Tog15,Bra15,Arn17,Arn18}
were performed on even-even Cr isotopes during the last two decades. Majority of the theoretical studies were based on SM  calculations. More recently  $^{46,48,50}$Cr isotopes~\cite{Nabi19,Ull20}
were investigated, using the pn-QRPA model, to study the effect of the deformation parameter on the calculated GT
strength distributions and ECC. In this work, even-even Cr nuclei along the $Z=28$ isotopic
chain were investigated in detail to study the deformation effect on the ECC rates and also
to further the calculation of their nuclear structure properties. It is to noted that
some results of $^{46,48,50}$Cr isotopes,  reported earlier in Refs.~\cite{Nabi19,Ull20}, were
added to this study to maintain systematics along the given isotopic chain.

The fundamental theme of the current work was to study the nuclear structure properties of even-even
$^{46-66}$Cr isotopes and to explore the effect of nuclear deformation parameters on the calculated
ECC of Cr isotopes. The energy levels and B(E2) values of the nuclei were computed within
the IBM-1 formalism. The geometric shapes of the isotopes were predicted by potting the contour plot
of the energy surface to obtain the deformation parameter. These isotopes display shape changing along
the isotopic chain,  $^{46,56,58,60}$Cr have spherical shape and $^{48-54,62-66}$Cr isotopes are the
axially deformed ones. $^{54}$Cr is closer to X(5) critical point, this isotope can be investigated in
detail. Moreover deformation parameters for each isotope were calculated using other theoretical models
and experimental data  as discussed in previous section.
To study the effect of nuclear deformation on calculated total GT strength and ECC, we selected eleven even-
even Cr isotopes ($^{46-66}$Cr). We computed ECC on the selected isotopes at three different stellar
temperatures T= (0.5 --1.5) MeV. The total GT strength values from the computed GT strength
distribution were calculated. The computed GT strength were widely dispersed among the daughter states
and they obeyed the ISR. The total GT strength were noted to have inverse relation with the deformation
parameter i.e with decrease in the deformation value, the total GT strength increased and vice versa.
The computed ECC increased with stellar temperature (0.5, 1.0 and 1.5) MeV.  Comparing the computed ECC
at different deformation values, it was found that they decreased with decrease in the value of
deformation parameter. We are in a process of studying a broader suite of nuclei to further probe the correlation effect of calculated ECC on nuclear deformation.

\section*{Acknowledgments}

J.-U. Nabi would like to acknowledge the support of the Higher
Education Commission Pakistan
through project numbers 5557/KPK
/NRPU/R$\&$D/HEC/2016, 9-5(Ph-1-MG-7)/PAK-TURK
/R$\&$D/HEC/2017 and Pakistan Science Foundation through project
number PSF-TUBITAK/KP-GIKI (02). 
M.  B\"{o}y\"{u}kata acknowledges
the support of the Scientific and Technical Research
Council of Turkey (T\"{U}B\.{I}TAK), under the project number 119T127.

\newpage

\begin{table}
\caption{The set of the parameters, in units of keV, for the model Hamiltonian in Eq. (\ref{ham}).
Here $N$ is the number of bosons, $\overline{\chi}$ is dimensionless and $\sigma$ is defined by Eq.
(\ref{rms})}
\label{par} \centering
\begin{tabular}{ccccccc}
\hline
 &~~$N$~~~~&$\epsilon_{d}$&~~$a_{1}$&~~$a_{2}$&~~$\overline{\chi}$&~~~~$\sigma$\\
\hline
$^{46}$Cr&~~~3~~~~&~~~~874,2~~~&~~~-48,5~&~~~~~-~~~~~&~~~-0,45~~~&~~~~~~6~~~\\
$^{48}$Cr&~~~4~~~~&~~1.515,5~~~&~~-233,3~&~~~-68,7~~~&~~~-1,02~~~&~~~~~24~~~\\
$^{50}$Cr&~~~3~~~~&~~1.200,2~~~&~~-156,5~&~~~-38,6~~~&~~~~-0,5~~~&~~~~~~8~~~\\
$^{54}$Cr&~~~3~~~~&~~~~802,4~~~&~~-152,5~&~~~-30,1~~~&~~~~-0,7~~~&~~~~~30~~~\\
$^{56}$Cr&~~~4~~~~&~~~~642,1~~~&~~~-53,6~&~~~~60,6~~&~~~~-0,6~~~&~~~~~~3~~~\\
$^{58}$Cr&~~~5~~~~&~~1.157,0~~~&~~~-46,4~&~~~~~-~~~~~&~~~~-1,1~~~&~~~~~~4~~~\\
$^{60}$Cr&~~~6~~~~&~~~~847,8~~~&~~~-33,5~&~~~~~-~~~~~&~~~~-0,5~~~&~~~~~17~~~\\
$^{62}$Cr&~~~7~~~~&~~~~777,4~~~&~~~-53,4~&~~~~~-~~~~~&~~~~-0,5~~~&~~~~~~0~~~\\
$^{64}$Cr&~~~7~~~~&~~~~733,6~~~&~~~-63,9~&~~~~~-~~~~~&~~~~-0,5~~~&~~~~~~0~~~\\
$^{66}$Cr&~~~6~~~~&~~~~502,3~~~&~~~-77,7~&~~~~~-~~~~~&~~~~-0,5~~~&~~~~~~0~~~\\
\hline
\end{tabular}
\end{table}
\newpage
\begin{table}
\caption{The experimental \cite{NNDC16} and calculated B(E2)$\downarrow$ values in units
of $10^{-2}$ $e^{2}b^{2}$ for the selected Cr isotopes. $\sigma$ in last column is defined by Eq.
(\ref{rms})} \label{be2} \centering
\begin{tabular}{@{}llllll}
\hline
~~~~~~~~~~~&~$J^\pi_i\rightarrow J^\pi_f$~~&~~~~~~EXP~~~~~~&~IBM-1~~&~~~~$e_b$~~&~~~~$\sigma$~~\\
\hline
~$^{46}$Cr~&~$2^+_{1}\rightarrow0^+_{1}$~~&~~1.86 (\emph{0.39})~~&~~1.84~~~~&~~~~0.072~~&~~~~0~~\\
\hline
~~~~~~~~~~~&~$2^+_{1}\rightarrow0^+_{1}$~~&~~3.21 (\emph{0.41})~~&~~2.96~~~~&~~~~~~~~~~~&~~~~~~~~~~~\\
~$^{48}$Cr~&~$4^+_{1}\rightarrow2^+_{1}$~~&~~2.80 (\emph{0.31})~~&~~3.92~~~~&~~~~0.065~~&~~~~6~~~~~~\\
~~~~~~~~~~~&~$6^+_{1}\rightarrow4^+_{1}$~~&~~3.01 (\emph{0.83})~~&~~3.43~~~~&~~~~~~~~~~~&~~~~~~~~~~~\\
~~~~~~~~~~~&~$8^+_{1}\rightarrow6^+_{1}$~~&~~2.49 (\emph{0.73})~~&~~2.08~~~~&~~~~~~~~~~~&~~~~~~~~~~~\\
\hline
~~~~~~~~~~~&~$2^+_{1}\rightarrow0^+_{1}$~~&~~2.08 (\emph{0.33})~~&~~1.74~~~~&~~~~~~~~~~~&~~~~~~~~~~~\\
~~~~~~~~~~~&~$4^+_{1}\rightarrow2^+_{1}$~~&~~1.60 (\emph{0.18})~~&~~2.36~~~~&~~~~~~~~~~~&~~~~~~~~~~~\\
~$^{50}$Cr~&~$6^+_{1}\rightarrow4^+_{1}$~~&~~2.41 (\emph{0.55})~~&~~2.31~~~~&~~~~0.045~~&~~~~7~~~~~~\\
~~~~~~~~~~~&~$2^+_{2}\rightarrow2^+_{1}$~~&~~0.01 (\emph{$^{+0.04}_{-0,01}$})~~&~~1.18~~~~&~~~~&~~~~\\
~~~~~~~~~~~&~$2^+_{2}\rightarrow0^+_{1}$~~&~~0.23 (\emph{0.08})~~&~~0.04~~~~&~~~~~~~~~~~&~~~~~~~~~~~\\
\hline
~~~~~~~~~~~&~$2^+_{1}\rightarrow0^+_{1}$~~&~~1.75 (\emph{0.07})~~&~~2.26~~~~&~~~~~~~~~~~&~~~~~~~~~~~\\
~~~~~~~~~~~&~$4^+_{1}\rightarrow2^+_{1}$~~&~~3.15 (\emph{1.09})~~&~~2.69~~~~&~~~~~~~~~~~&~~~~~~~~~~~\\
~$^{54}$Cr~&~$6^+_{1}\rightarrow4^+_{1}$~~&~~2.18 (\emph{0.61})~~&~~1.80~~~~&~~~~0.074~~&~~~~4~~~~~~\\
~~~~~~~~~~~&~$2^+_{2}\rightarrow2^+_{1}$~~&~~0.85 (\emph{0.48})~~&~~1.29~~~~&~~~~~~~~~~~&~~~~~~~~~~~\\
~~~~~~~~~~~&~$2^+_{2}\rightarrow0^+_{1}$~~&~~0.02 (\emph{0.001})~~&~~0.03~~~&~~~~~~~~~~~&~~~~~~~~~~~\\
~~~~~~~~~~~&~$0^+_{2}\rightarrow2^+_{1}$~~&~~1.21 (\emph{$^{+0.36}_{-0,48}$})~~&~~0.63~~~~&~~~~&~~~~\\
\hline
~$^{56}$Cr~&~$2^+_{1}\rightarrow0^+_{1}$~~&~~1.11 (\emph{0.38})~~&~~1.12~~~~&~~~~0.045~&~~~~0~~~~~~\\
\hline
~$^{58}$Cr~&~$2^+_{1}\rightarrow0^+_{1}$~~&~~1.97 (\emph{0.56})~~&~~1.97~~~~~&~~~~0.053~&~~~~0~~~~~~\\
\hline
~$^{60}$Cr~&~$2^+_{1}\rightarrow0^+_{1}$~~&~~2.21 (\emph{0.29})~~&~~2.18~~~~~&~~~~0.050~&~~~~0~~~~~~\\
\hline
~$^{62}$Cr~&~$2^+_{1}\rightarrow0^+_{1}$~~&~~3.25 (\emph{0.44})~~&~~3.28~~~~~&~~~~0.049~&~~~~0~~~~~~\\
\hline
~$^{64}$Cr~&~$2^+_{1}\rightarrow0^+_{1}$~~&~~3.19 (\emph{0.08})~~&~~3.20~~~~~&~~~~0.047~&~~~~0~~~~~~\\
\hline
\end{tabular}
\end{table}
\newpage
\begin{table}[ht]
	\caption{The calculated total GT strength values
in the EC direction for the selected Cr isotopes as a function of
deformation parameters.}
	\centering
	\begin{tabular}{c c c c }
		\hline\hline
	    Isotopes & Deformation parameters & $\beta$ & $\sum B(GT_+)$  \\ [0.5ex]
		\hline
		          & $\beta_{Mac-mic}~$   & 0.028& 9.13\\
		$^{46}$Cr & $\beta_{IBM-1}~~~$   & 0.000  & 9.67 \\
		          & $\beta_{GM}~~~~~~~$  & 0.000  & 9.67 \\
		          & $\beta_{B(E2)}~~~~~$ & 0.288 & 8.97\\
		\hline
		          & $\beta_{Mac-mic}~$   & 0.236& 8.94\\
		$^{48}$Cr & $\beta_{IBM-1}~~~$   & 0.760  & 8.46 \\
		          & $\beta_{GM}~~~~~~~$  & 0.299  & 8.77 \\
		          & $\beta_{B(E2)}~~~~~$ & 0.368 & 8.63\\
		\hline
		          & $\beta_{Mac-mic}~$   & 0.141& 8.58\\
		$^{50}$Cr & $\beta_{IBM-1}~~~$   & 0.586  & 7.22\\
		          & $\beta_{GM}~~~~~~~$  & 0.277  & 7.73 \\
		          & $\beta_{B(E2)}~~~~~$ & 0.290  & 7.66\\
		\hline
		$^{52}$Cr & $\beta_{Mac-mic}~$   & 0.023& 8.48\\
		          & $\beta_{B(E2)}~~~~~$ & 0.223  & 7.77\\
		\hline
		          & $\beta_{Mac-mic}~$   & 0.134& 8.50\\
		$^{54}$Cr & $\beta_{IBM-1}~~~$   & 0.362  & 7.23\\
		          & $\beta_{GM}~~~~~~~$  & 0.095  & 8.91 \\
		          & $\beta_{B(E2)}~~~~~$ & 0.251  & 7.75\\
		\hline
		          & $\beta_{Mac-mic}~$   & 0.163& 3.48\\
		$^{56}$Cr & $\beta_{IBM-1}~~~$   & 0.000  & 4.21\\
		          & $\beta_{GM}~~~~~~~$  & 0.000  & 4.21\\
		          & $\beta_{B(E2)}~~~~~$ & 0.195  & 3.25\\
		\hline
		          & $\beta_{Mac-mic}~$   & 0.164& 2.48\\
		$^{58}$Cr & $\beta_{IBM-1}~~~$   & 0.000  & 3.19 \\
		          & $\beta_{GM}~~~~~~~$  & 0.000  & 3.19\\
		          & $\beta_{B(E2)}~~~~~$ & 0.254  & 2.11\\
		\hline
		          & $\beta_{Mac-mic}~$   & 0.151& 1.77\\
		$^{60}$Cr & $\beta_{IBM-1}~~~$   & 0.000  & 2.18\\
		          & $\beta_{GM}~~~~~~~$  & 0.000  & 2.18\\
		          & $\beta_{B(E2)}~~~~~$ & 0.263  & 1.56\\
		\hline
		          & $\beta_{Mac-mic}~$   & 0.081& 1.30\\
	    $^{62}$Cr & $\beta_{IBM-1}~~~$   & 0.492  & 1.13 \\
		          & $\beta_{GM}~~~~~~~$  & 0.262  & 1.28 \\
		          & $\beta_{B(E2)}~~~~~$ & 0.312  & 1.22\\
		\hline
		          & $\beta_{Mac-mic}~$   & 0.023& 1.09\\
		$^{64}$Cr & $\beta_{IBM-1}~~~$   & 0.601  & 0.70\\
		          & $\beta_{GM}~~~~~~~$  & 0.355  & 0.81 \\
		          & $\beta_{B(E2)}~~~~~$ & 0.303  & 0.85\\
		\hline
		          & $\beta_{Mac-mic}~$   & 0.009& 0.89\\
		$^{66}$Cr & $\beta_{IBM-1}~~~$   & 0.701  & 0.64\\
		          & $\beta_{GM}~~~~~~~$  & 0.451  & 0.71\\
		\hline
	\end{tabular}
	\label{A}
\end{table}

\newpage

\newpage

\newpage
\begin{figure}
	\centering
	\includegraphics[height=7in, width=7in]{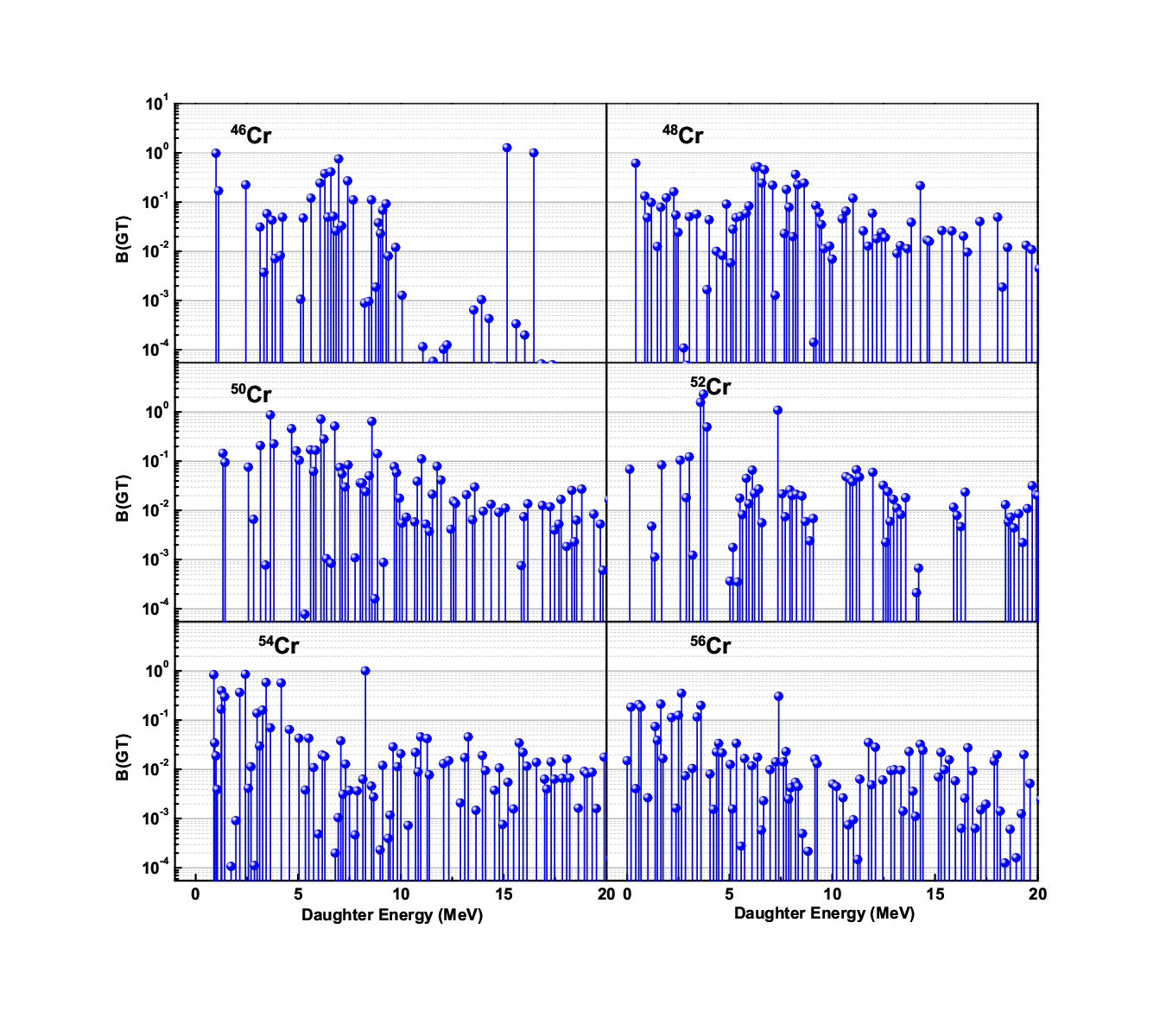}
	\caption{The pn-QRPA computed GT strength distributions along electron capture direction for $^{46,48,50,52,54,56}$Cr as a function of excitation energy in daughter.}
	\label{46-56CR_GT}
\end{figure}\label{46-56CR_GT}

\newpage
\begin{figure}
	\centering
	\includegraphics[height=7in, width=7in]{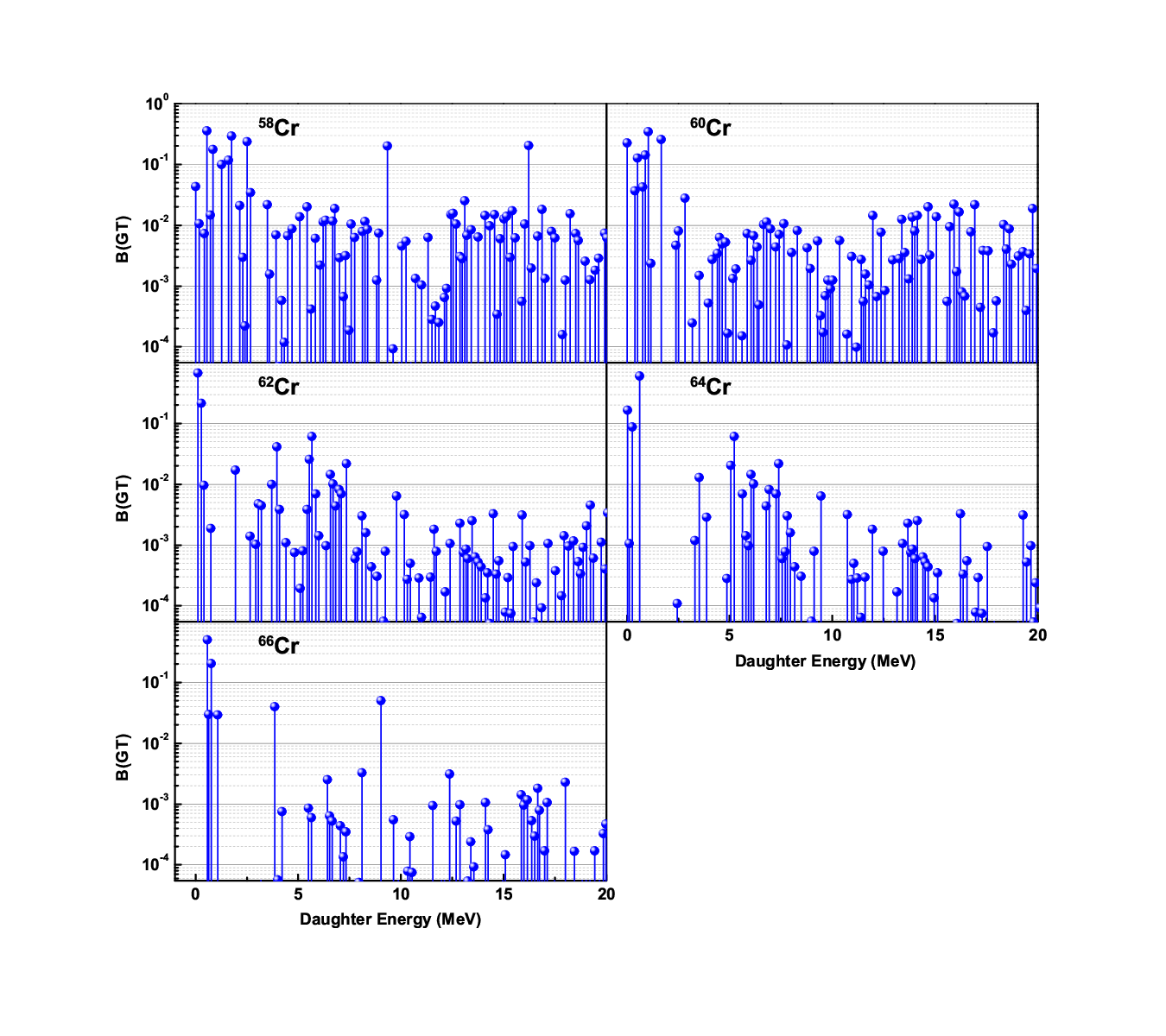}
	\caption{Same as Fig. \ref{46-56CR_GT}, but for $^{58,60,62,64,66}$Cr.}
	\label{58-66CR_GT}
\end{figure}\label{58-66CR_GT}
\begin{figure}
\begin{center}
	\includegraphics[width=28cm]{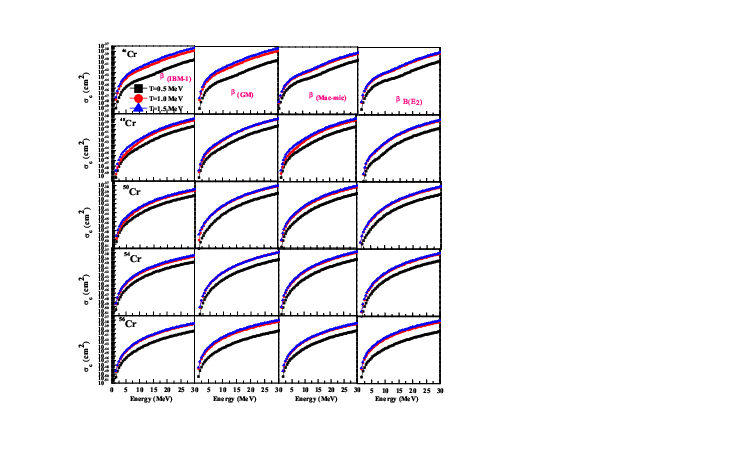}
\end{center}
	\caption{ Calculated ECC on $^{46,48,50,54,56}$Cr as a function of incident electron energy
             for different values of deformation parameter and stellar temperature.}
\label{46-64CR_a}
\end{figure}\label{46-64CR_a}

\newpage
\begin{figure}
	\centering
	\includegraphics[height=8in, width=9in]{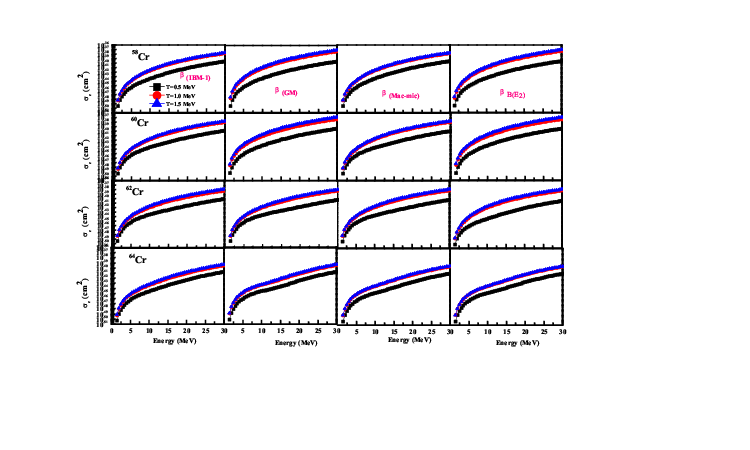}
	\caption{ Same as Fig. \ref{46-64CR_a}, but for $^{58,60,62,64}$Cr.}
\label{46-64CR_b}
\end{figure}\label{46-64CR_b}

\newpage
\begin{figure}
	\centering
	\includegraphics[height=7in, width=7in]{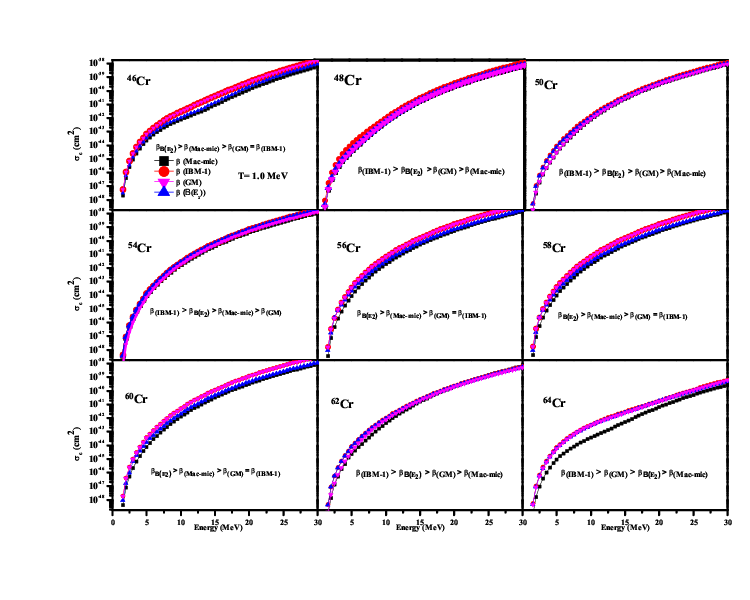}
	\caption{ Comparison of calculated ECC on $^{46,48,50,54,56,58,60,62,64}$Cr with different $\beta$ values at a selected temperature of 1.0 MeV.}
\label{d46-64CR}
\end{figure}\label{d46-64CR}

\end{document}